\documentstyle[12pt,twoside,psfig]{article}

\def\be{\begin{equation}}
\def\ee{\end{equation}}
\def\bea{\begin{eqnarray}}
\def\ena{\end{eqnarray}}

\def\a{\alpha}
\def\b{\beta}

\def\d{\delta}
\def\e{\epsilon}           

\def\m{\mu}
\def\n{\nu}

\def\p{\pi}                
\def\q{\theta}                    
\def\r{\rho}                      
\def\s{\sigma}                    
\def\t{\tau}

\def\G{\Gamma}

\def\del{\partial}

\def\co{{\cal O}}

\catcode`@=11

\def\un#1{\relax\ifmmode\@@underline#1\else
$\@@underline{\hbox{#1}}$\relax\fi}

{\catcode`\'=\active \gdef'{{}^\bgroup\prim@s}}

\def\magstep#1{\ifcase#1 \@m\or 1200\or 1440\or 1728\or 2074\or 2488\or
       2986\fi\relax}   

\catcode`@=12

\def\bop#1{\setbox0=\hbox{$#1M$}\mkern1.5mu
	\vbox{\hrule height0pt depth.04\ht0
	\hbox{\vrule width.04\ht0 height.9\ht0 \kern.9\ht0
	\vrule width.04\ht0}\hrule height.04\ht0}\mkern1.5mu}

\def\leftrightarrowfill{$\mathsurround=0pt \mathord\leftarrow \mkern-6mu
       \cleaders\hbox{$\mkern-2mu \mathord- \mkern-2mu$}\hfill
       \mkern-6mu \mathord\rightarrow$}
\def\dvec#1{\vbox{\ialign{##\crcr
       \leftrightarrowfill\crcr\noalign{\kern-1pt\nointerlineskip}
       $\hfil\displaystyle{#1}\hfil$\crcr}}}          
\def\hook#1{{\vrule height#1pt width0.4pt depth0pt}}
\def\leftrighthookfill#1{$\mathsurround=0pt \mathord\hook#1
       \hrulefill\mathord\hook#1$}
\def\underhook#1{\vtop{\ialign{##\crcr                 
       $\hfil\displaystyle{#1}\hfil$\crcr
       \noalign{\kern-1pt\nointerlineskip\vskip2pt}
       \leftrighthookfill5\crcr}}}
\def\smallunderhook#1{\vtop{\ialign{##\crcr      
       $\hfil\scriptstyle{#1}\hfil$\crcr
       \noalign{\kern-1pt\nointerlineskip\vskip2pt}
       \leftrighthookfill3\crcr}}}

\def\sfrac#1#2{{\vphantom1\smash{\lower.5ex\hbox{\small$#1$}}\over
       \vphantom1\smash{\raise.4ex\hbox{\small$#2$}}}} 
\def\bfrac#1#2{{\vphantom1\smash{\lower.5ex\hbox{$#1$}}\over
       \vphantom1\smash{\raise.3ex\hbox{$#2$}}}}      
\def\afrac#1#2{{\vphantom1\smash{\lower.5ex\hbox{$#1$}}\over#2}}  
\def\on#1#2{{\buildrel{\mkern2.5mu#1\mkern-2.5mu}\over{#2}}}
\def\ddt#1{\on{\hbox{\LARGE .\kern-2pt.}}#1}             
\def\tdt#1{\on{\hbox{\LARGE .\kern-2pt.\kern-2pt.}}#1}   

\def\boxes#1{
       \newcount\num
       \num=1
       \newdimen\downsy
       \downsy=-1.5ex
       \mskip-2.8mu
       \bo
       \loop
       \ifnum\num<#1
       \llap{\raise\num\downsy\hbox{$\bo$}}
       \advance\num by1
       \repeat}
\def\boxup#1#2{\newcount\numup
       \numup=#1
       \advance\numup by-1
       \newdimen\upsy
       \upsy=.75ex
       \mskip2.8mu
       \raise\numup\upsy\hbox{$#2$}}

\newskip\humongous \humongous=0pt plus 1000pt minus 1000pt

\newif\ifdtup

\parskip=\medskipamount  
\def\baselinestretch{1.2}
\thicklines              
\def\border{
 }
\def\headpic{
 }
\def\title#1#2#3#4{\begin{document}
       \border
       \headpic
       {\hbox to\hsize{#4 \hfill ITP-SB-#3}}\par
       \begin{center}\vskip.8in minus.1in
       {\Large\bf #1}\\[.5in minus.2in]{#2}
       \vskip1.4in minus1.2in {\bf ABSTRACT}\\[.1in]\end{center}
       \begin{quotation}\par}
\def\author#1#2{#1\\[.1in]{\it #2}\\[.1in]}
\def\ITP{\footnote{Work supported by National Science Foundation
  grant PHY 89-08495.}\\[.1in] {\it Institute for Theoretical Physics\\
  State University of New York, Stony Brook, NY 11794-3840}\\[.1in]}
\def\endtitle{\par\end{quotation}\vskip3.5in minus2.3in\newpage}

\def\camera#1#2{
       \topmargin=.46in
       \textheight=22cm
       \textwidth=15cm
       \hsize=15cm
       \oddsidemargin=.28in
       \evensidemargin=.28in
       \marginparsep=0in
       \parindent=1.15cm
       \pagestyle{empty}
       \def\rm{\sf}
       \begin{document}
       \begin{center}{\Large\bf #1}\\[.5in minus.2in]{\bf #2}
       \vskip1in minus.8in {ABSTRACT}\\[.1in]\end{center}
       \renewcommand{\baselinestretch}{1}\small\normalsize
       \begin{quotation}\par}
\def\endabstract{\par\end{quotation}
       \renewcommand{\baselinestretch}{1.2}\small\normalsize}

\def\xpar{\par}                                       
\def\header{
 }
\def\letterhead{
 \header
 \font\sflarge=helvetica at 14pt   
 \leftskip=2.8in\noindent\phantom m\\[-.54in]
       {\large\sflarge STATE UNIVERSITY OF NEW YORK}
       {\scriptsize\sf INSTITUTE FOR THEORETICAL PHYSICS\\[-.07in]
       STONY BROOK, NY 11794-3840\\[-.07in]
       Tel: (516) 632-7979}
 \vskip.3in\leftskip=0in}
\def\letterneck#1#2{\par{\hbox to\hsize{\hfil {#1}\hskip 30pt}}\par
       \begin{flushleft}{#2}\end{flushleft}}
\def\letterhat{\parskip=\bigskipamount \def\baselinestretch{1}}
\def\head#1#2{\letterhat\begin{document}\letterhead\letterneck{#1}{#2}}
\def\multihead#1#2{\thispagestyle{empty}\setcounter{page}{1}
       \letterhead\letterneck{#1}{#2}}        
\def\shead#1#2{\letterhat\begin{document}\sletterhead
       \letterneck{#1}{#2}}
\def\multishead#1#2{\thispagestyle{empty}\setcounter{page}{1}
       \sletterhead\letterneck{#1}{#2}}
\def\multisig#1{\goodbreak\bigskip{\hbox to\hsize{\hfil Kind regards,
       \hskip 30pt}}\nobreak\vskip .5in\begin{quote}\raggedleft{#1}
       \end{quote}}
\def\sig#1{\multisig{#1}\end{document}}

\def\watch{
 \newcount\hrs
 \newcount\mins
 \newcount\merid
 \newcount\hrmins
 \newcount\hrmerid
 \hrs=\time
 \mins=\time
 \divide\hrs by 60
 \merid=\hrs
 \hrmins=\hrs
 \divide\merid by 12
 \hrmerid=\merid
 \multiply\hrmerid by 12
 \advance\hrs by -\hrmerid
 \ifnum\hrs=0\hrs=12\fi
 \multiply\hrmins by 60
 \advance\mins by -\hrmins
 \number\hrs:\ifnum\mins<10 {0}\fi\number\mins\space\ifnum\merid=0
 AM\else PM\fi}
\def\half{\frac{1}{2}}

\def\eq{\begin{equation}}
\def\eqe{\end{equation}}
\def\eqa{\begin{eqnarray}}
\def\eqae{\end{eqnarray}}

\def\be{\begin{equation}}
\def\ee{\end{equation}}
\def\bea{\begin{eqnarray}}
\def\ena{\end{eqnarray}}

\def\to{\rightarrow}

\begin{document}

\thispagestyle{empty}
\begin{flushright}
{\sc ITP-SB}-95-12
\end{flushright}
\vspace{1cm}
\setcounter{footnote}{0}
\begin{center}
{\LARGE\sc{Loop calculations in quantum-mechanical non-linear sigma
 models\footnote{This research was supported in part by NSF grant no
 PHY9309888.}
    }}\\[14mm]

\sc{Jan de Boer\footnote{e-mail: deboer@insti.physics.sunysb.edu},
    Bas Peeters\footnote{e-mail: peeters@insti.physics.sunysb.edu},
    Kostas Skenderis\footnote{e-mail: kostas@insti.physics.sunysb.edu} and
    Peter van Nieuwenhuizen\footnote{e-mail:
                          vannieu@insti.physics.sunysb.edu},}\\[5mm]
{\it Institute for Theoretical Physics\\
State University of New York at Stony Brook\\
Stony Brook, NY 11794-3840, USA}\\[20mm]

{\sc Abstract}\\[2mm]
\end{center}

\noindent
By carefully analyzing the relations between operator methods and the
discretized and continuum path integral formulations of
quantum-mechanical systems, we have
found the correct Feynman rules
for one-dimensional path integrals in curved spacetime.
Although the prescription
how to deal with the products of distributions that appear in the
computation of Feynman diagrams in configuration space is
surprising, this prescription follows unambiguously from
the discretized path integral. We check our results by an explicit
two-loop calculation.

\vfill

\newpage



\baselineskip=24pt

One-dimensional path integrals are interesting for a variety of reasons.
They are the simplest toy models of higher dimensional path integrals,
and since they describe quantum mechanics, which is a finite theory,
it should be possible to understand them completely and rigorously. Indeed,
it is relatively straightforward to reproduce some standard quantum mechanical
facts for the harmonic oscillator
and other simple systems
from the corresponding path integrals. However,
to do the same for a point particle in curved space turns out to be a much
more delicate problem \cite{witt,sch}.
The corresponding path integrals can be used to
compute one-loop anomalies of $n$-dimensional quantum field theories.
This was first shown by Alvarez-Gaum\'e and Witten for various chiral
anomalies \cite{ref1}, and later for trace anomalies in \cite{ref2}. In the
first case, only one-loop worldline graphs contribute, but in the second
case one needs to go to higher loops, and details of the action and
measure can no longer be neglected. This is yet another reason to
properly understand one-dimensional path integrals. Inspired by this
application we will in this article consider the path integral for
a point particle in curved space, defined on a finite time interval
$[-\beta,0]$. Since the corresponding Hamiltonian $\hat{H}$
appears as a regulator
when we write anomalies $A$ as the regularized trace of a Jacobian $\hat{J}$,
$A={\rm Tr}(\hat{J} \exp(-\beta \hat{H} ))$, the Hamiltonian and its
operator ordering are completely fixed by the properties of the original
quantum field theory. Therefore, we will in this article
only focus on the path integral
computation of the transition element\footnote{We work
in Euclidean space. For loop calculations, there is no essential difference
with Minkowski space, and we ignore instantons, since we are interested in
the short time expansion of $T$ only.}
 $T=<z|\exp(-\beta \hat{H})|y>$,
given a fixed quantum Hamiltonian $\hat{H}$. The supersymmetric
generalization, a full-fledged path integral treatment of (Majorana)
fermions and internal symmetry
ghosts and the explicit evaluation of various anomalies
will be discussed in a future publication \cite{ref3}.

At first sight it may seem surprising that one can find a fundamental
new result in such an arcane subject as path integrals for a point
particle. However, most of the vast literature on this subject deals
with infinite intervals, and those articles in which finite intervals
are considered do not contain explicit higher loop calculations. Finally,
some authors do not require that path integrals should reproduce the
results obtained from a Hamiltonian approach, but rather they
freely invent a new definition of Feynman path integrals directly in
configuration space. Because of the connection with anomalies, we define
path integrals as derived from the Hamiltonian approach. In \cite{bas},
an explicit complete two-loop calculation of the trace anomaly was
performed, using mode regularization to define products of $\delta$-
and $\theta$-functions. Surprisingly, the results disagreed with the results
of \cite{ref2} obtained from the heat kernel. A thorough and lengthy study
finally revealed that mode regularization is incompatible with the
discretized Hamiltonian approach.

\begin{figure}
\centerline{\hbox{\psfig{figure=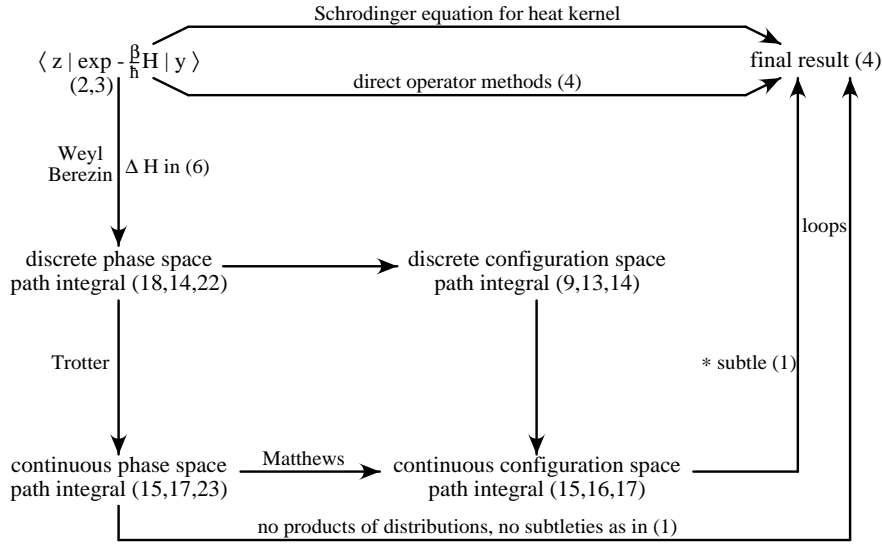,width=5in}}}
\caption{Flow chart with equations. The asterisk locates the
 main result.}
\end{figure}

Various ways to obtain an explicit expression for $T$ have been depicted in
figure 1. Our main result, denoted by a star in figure 1, will be
the precise rules how
to obtain $T$ at the perturbative level
from a continuous configuration space path integral.
Our strategy will be to insert complete sets of states in
$<z|\exp(-\beta \hat{H})|y>$, leading to a discrete phase space path integral.
As was noted by Berezin \cite{ber}, the classical
Hamiltonian that appears in the
phase space path integral is related to $\hat{H}$ via Weyl ordering
(explained below, see
(\ref{path7}),(\ref{path8})),
and is not covariant\footnote{In our subsequent discussion of
non-covariance we mean of course covariance in target space, not to be
confused with problems with Lorentz invariance on the world-sheet,
which can exist only in dimensions
larger than one.}. In \cite{geji} it was
shown that the non-covariant terms in the Hamiltonian are
needed for consistency by doing a two-loop calculation for the continuous
phase space path integral
for a free point particle submitted to an arbitrary co-ordinate
transformation.
How can one understand this non-covariant term,
which also shows up in the configuration space path integral? After all,
the final result for the transition element is covariant. The answer
is that path integrals are defined as the limit of time-discretized path
integrals, and the latter use the midpoint rule,
i.e. they replace in a small time interval
the co-ordinates in the Hamiltonian
or Lagrangian by the average of their values
at the endpoints of the interval. Clearly, taking the average of the
co-ordinates of two
points in curved space-time is not a covariant operation. This non-covariance
conspires with the non-covariant term in the action to yield a covariant
result for $T$. After taking the continuum limit, one does
not see the midpoint rule anymore, nor the fact that it is non-covariant.
The reason one still finds a covariant answer for $T$ from the
continuum path integrals lies in the Feynman rules, and the way
in which we compute the Feynman diagrams. From the discretized path integral,
we not only find the continuum action and vertices
and propagators, but also the prescription how to
compute Feynman diagrams. These rules and their derivation are new, and
constitute our main result, indicated by an asterisk in figure 1.
The unambiguous answer will (not surprisingly) look
a lot like lattice regularization, and this non-covariant regularization
is the remnant of the non-covariant midpoint rule.
Only with this
regularization can one reproduce the covariant answer for $T$.
By explicit calculation we have found that
more conventional methods such as momentum cut-off or
mode-regulatization \cite{ref2}
do not yield the correct answer for $T$ \cite{ref3,bas}!
In \cite{witt}, DeWitt defines the Feynman path integral from
covariance arguments, without recourse to the Hamiltonian formalism,
and does not find the non-covariant term (which is proportional
to the product of two Christoffel symbols) in his action.
Since he works in an inertial frame, he finds no contradictions at the
2-loop level, but our results demonstrate that from 3 loops on, the
$\Gamma\Gamma$ term, or something equivalent, can not be neglected.

The non-covariant time-discretized path integral
which we use in this article
follows naturally
if one inserts complete sets of $x$- and $p$-eigenstates
and only retains terms of order $1/N$ in the phase space path integral.
There are possibilities
to construct a covariant time-discretized
path integral, but these are all very
problematic. One can try to construct a path integral which is
discretized using a mid-geodesic rule, but we do not know to what
operator ordering this corresponds, and neither what the Feynman rules
should be. Another possibility is to glue
together $N$ times the exact result for $T$ with
$\beta$ replaced by $\beta/N$, keeping only terms up to order $1/N$,
and to take this as the definition of the discretized path integral.
This will in the limit $N\rightarrow \infty$ certainly reproduce the
correct answer for $T$, but the corresponding continuum configuration
space path integral
turns out to have a non-local kernel (see (\ref{path6})),
and it is not clear how
to derive the Feynman rules. Hence, we will stick to the local
action in the path
integral with its non-covariant term. It is an
interesting phenomenon that one can obtain the same result for $T$
by taking the large $N$ limit of the $N^{\rm th}$ power of two different
kernels that even differ by terms of order $1/N$.

The path integrals we consider are defined on a finite time interval
$[-\beta,0]$. Consequently, propagators depend on $\s$ and $\t$, and not
only on $\s-\t$, so that the usual momentum space manipulations of Feynman
cannot be used to give meaning to the product of distributions which always
occur in Feynman graphs. We shall therefore work in co-ordinate space.
Although individual graphs are divergent by power counting, all infinities
cancel if one introduces the ghosts of ref \cite{ref2}.
Consequently, there is no need for renormalization. However, we have
to specify how to deal with the products of distributions that appear.
For example, the discretized path integral tells us unambiguously
that\footnote{Such a rule has appeared previously in the literature
in e.g. \cite{witt,conn,alfdam}. The rules $\delta(t,t)dt=1$ and
$\theta(t,t)=1/2$ are proposed in \cite{witt}, page 333, as `two
formal identifications' and combined they yield (\ref{path1}).
In \cite{conn,alfdam} similar rules were found necessary to reproduce
the non-covariant term in the action from respectively an auxiliary
fermionic and an auxiliary ghost system. We give a proper derivation of
these and other rules.}
\eq
\int^0_{-1} \int^0_{-1} \d (\s-\t) \q (\s-\t) \q (\t-\s)  = {1\over 4}.
\label{path1}
\eqe
Mathematically, products of distributions are not well-defined. If we
expand $\d$ and $\q$ in a complete set of
trigonometric functions then the left hand
side of (\ref{path1}) yields a not absolutely convergent sum. Indeed,
one might as well claim it is zero
(arguing that $\q (\s-\t) \q
(\t-\s)$ is a bounded function which is almost everywhere zero) or 1/3
(arguing that $\d (\s-\t)=\del_\s \q (\s-\t) $
or using mode regularization). Besides (\ref{path1}),
the discretized path integral also yields the equal-time contractions
of two-point functions, which we need but which in
continuum path integrals are often claimed to be undefined \cite{ref4}.
We now give details.

For a quantum field theory with scalar fields coupled to external
gravity, the consistent regulator can be found from a general method
\cite{ref5},
and reads $g^{-1/4} \del_i g^{1/2} g^{ij} \del_j g^{-1/4}$ where $g
= \det g_{ij}$.  Following \cite{ref1}, we represent this regulator as the
Hamiltonian of the nonlinear sigma model with $S = \int^0_{-\b}
{1\over 2} g_{ij} \dot{x}^i \dot{x}^j dt$, but with the same operator
ordering as the regulator
\eq
\hat{H} = {1\over 2} g^{-1/4} p_i \sqrt{g} g^{ij} p_j g^{-1/4}
\label{path2}
\eqe
This operator is Einstein (general co-ordinate) invariant \cite{witt}.
Using complete sets of position and momentum eigenstates $( \int | x >
\sqrt{g(x)} < x | d^n x = \int | p >< p| d^n p=I)$, we can evaluate the
transition element
\eq
T = < z | e^{-{\b \hat{H} \over \hbar}}  | y > = \int < z | e^{-{\b
\hat{H} \over
\hbar}} | p > < p|y> d^n p
\label{path3}
\eqe
by expanding the exponent and moving all operators $\hat{x}^j $ to
the left and $\hat{p}_j$ to the right, keeping track of commutators.
We stress that the answer for $T$ is thus finite and unambiguous.
To given order in $\b$
all the terms in the expansion of $\exp(-\b\hat{H}/\hbar)$ contribute, but
only a finite number of commutators in each
term are needed, and one may resum the infinite series.
(Each commutator removes one factor $p$, and integration over $p$
shows that $p$ is of order $z-y$. We take $z-y$ of order $\b^{1/2}$, which
is often justified by arguments concerning Brownian motion \cite{sch}).
To given order
in $\b$, $T$ is then a Gaussian factor involving $p$, multiplied by a
polynomial in $p$. The Gaussian $p$ integral can then easily be performed,
and in this way
we have found (after much algebra) the following result, exact through
order $\b$ (this result agrees with \cite{witt2},
see also \cite{gra,shie})
\eq
T=  (2 \p \hbar \b)^{-n/2} \left( \exp -
{1\over \hbar} S_{cl}^B \left[ z , y , \b \right] \right)  D \exp {-\b
\hbar \over 12} R
\label{path5}
\eqe
Here $(2\p \hbar \b)^{-n/2}$ is the usual Feynman factor, $S_{cl}^B
[z, y, \b]$ is the classical action for a geodesic with $x^j (-\b ) = y^j,
x^j (0) = z^j ,$ while $ -{1\over 12} R$ is the trace anomaly in $n=2$
dimensions, and $D$ contains the Van Vleck determinant which
gives the one-loop corrections to $T$
\eqa
D &=& \b^{n/2} g^{-1/4} (z) \det \left( - {\del \over \del z^i} {\del \over
\del y^j}  S_{cl} [z, y, \b] \right)^{\frac{1}{2}}
 g^{-1/4} (y) \nonumber\\
&=& 1 - {1\over 12} R_{ij} (z) (z-y)^i (z-y)^j + \co (\b^{3/2} )
\label{path6}
\eqae
Note that if we view $T$ for $\b\rightarrow 0$ as the kernel of
a continuum path integral action, then $D$ corresponds to a non-local
term.
Requiring that the composition of two transition elements $T(\b)$ reproduces
$T(2\b)$ up to order ${\cal O}(\b^{3/2})$, fixes the $R_{ij}$ term but
does not fix the coefficient of the $R$ term.
Clearly, all factors in (\ref{path5}) are Einstein invariant. By rescaling $t
= \b\t$, (\ref{path5}) only depends on the product $\hbar \b$, which is
the loop counting parameter.  We must now construct the path integral
whose loop expansion reproduces
 (\ref{path5}).

 We begin by {\it rewriting} $\hat{H}$ in Weyl-ordered form. Weyl ordering
\cite{ber,wey1,miz,sato,wey2}
is defined by $(n+m)! (p^nq^m)_W=\partial_a^n \partial_b^m (ap+bq)^{m+n}$,
and we find the well known result \cite{miz,omo2}
 \eq
 \hat{H} = ({1\over 2} g^{ij} p_i p_j )_{\stackrel{}{W} }
 + {\hbar^2 \over 8} \left(
\G^\r_{\m\s} \G^\s_{\n\r}  g^{\m\n} + R \right)
\label{path7}
\eqe
Then we use the fact that Weyl
ordering leads to the midpoint rule \cite{ber}
$$
\int <x_{k+1}  | G_W | p > < p | x_k > d^n p
 =
\int
 G (p, x_{k+1/2} ) < x_{k+1} | p >< p | x_k > d^n p,
$$
where we defined $x_{k+1/2}=(x_{k+1}+x_k)/2$, and $G$ is any function
of $p,q$. Applying this to $G=\exp(-\e H/\hbar)$ and using
$\exp(-\e (H)_W/\hbar)=\exp(-\e H/\hbar)_W+{\cal O}(\e^2)$ we find
\eqa
\int <x_{k+1}  | \exp - {\e \over \hbar} (H)_W | p > < p | x_k > d^n p
& =& \nonumber \\
& &
\!\!\!\!\!\!\!\!\!\!\!\!\!\!\!\!\!\!\!\!\!\!\!\!\!\!\!\!\!\!\!\!\!\!
\!\!\!\!\!\!\!\!\!\!\!\!\!\!\!\!\!\!\!\!\!\!\!\!\!\!\!\!\!\!\!\!\!\!
\int \exp -
{\e \over \hbar} H (p, x_{k+1/2} ) < x_{k+1} | p >< p | x_k > d^n p
+ {\cal O}(\e^2).
\label{path8}
\eqae

In the construction of the discretized path integral we will neglect the
terms of order ${\cal O}(\e^2)$. This is a subtle issue \cite{shie}. Naively,
it is based on the fact that $\lim_{N\rightarrow \infty}
(1+aN^{-1}+bN^{-2}+\ldots)^N=e^a$ is independent of $b$. For operators,
it usually goes under the name of the Trotter formula, although we are
not aware of a rigorous proof for the case of curved space.
One of the
subtleties is the meaning of ${\cal O}(\e^2)$. In (\ref{path8})
we mean that we view $p$ and $q$ of order one, but if we integrate
over $p$ and remember that $H$ contains a term proportional to
$\e p^2$ we could argue that $p$ is really of order $\e^{-1/2}$, by
looking at the identity $\int dp p^2 \exp(-\e p^2)/\int dp \exp(-\e p^2)=
1/(2\e)$. An identity like (\ref{path8}), valid up to order
${\cal O}(\e^{3/2})$ while counting $p$ as order $\e^{-1/2}$ can be written
down for (\ref{path3}),
and it contains additional terms on the right hand side \cite{bas,gra,belg}.
After performing the $p$ integrals we are left with extra terms
proportional to $x_{k+1}-x_k$, where this difference has order
$\e^{1/2}$. These terms cannot be obtained from the discretization of
a continuum action unless it contains non local terms.
It has been argued
in \cite{lasch} that they can be replaced by effective potentials
depending only on $x_{k+1/2}$. Now it is easy to see\footnote{Let
us sketch the argument. Since the extra terms
in (\ref{path8}) appear when we count $p$ as $\e^{-1/2}$,
they must contain $p$ and vanish as $p\rightarrow 0$. The effective
potentials associated to some function of $\dot{q}$ can be found
be performing an integral over $\dot{q}$. Denoting the extra terms
by ${\cal E}$ we find that (i), (ii) and (iii) combined yield
$\int d\dot{q} \int dp \exp(ip\dot{q}-H) {\cal E}$. Interchanging
the order of integration yields a delta function of $p$ and the
result follows from ${\cal E}(p=0)=0$.}
 that the combined
operation of (i) keeping extra terms in (\ref{path8}), (ii) doing
the $p$-integrals and (iii) replacing the extra terms by effective
potentials is exactly zero. This is ultimately the justification why
(\ref{path8}) is sufficient for our purposes. It follows that the
difference between the kernel in (\ref{path5}) and that in
(\ref{path8}) is exactly the difference between terms containing
$z-y$ (including the $R_{ij}$ term in (\ref{path6}))
and their effective potentials.  We believe that all
equivalent kernels are related by such manipulations \cite{belg}, although we
have no general proof of this.

Coming back to the derivation of $T$ from (\ref{path8}), we insert
$N-1$ sets of $x$-eigenstates and $N$ sets of
$p$-eigenstates into $< z  | \exp - \frac{\b}{\hbar} \hat{H} | y >$, and
we arrive at the discretized phase space path integral using
$< x | p > = (2 \p \hbar  )^{-n/2} (\exp {i\over \hbar} p \cdot
x) g^{-1/4} (x )$.
Integrating out the $N$ momenta we find the discretized configuration space
path
integral, with $N$ factors $g^{1/2}(x_{k+1/2})$ in the measure from the
$p$ integrals, $N$ products $g^{-1/4}(x_{k+1})g^{-1/4}(x_k)$ from the inner
products $<x|p>$ and $N-1$ factors $g^{1/2}(x_k)$ from the completeness
relation in $x$-space. The action is given by
\eq
S=\sum_{k=0}^{N-1} \frac{1}{2\epsilon} g_{ij}(x_{k+1/2})
 (x_{k+1}-x_k)^i (x_{k+1}-x_k)^j  - \frac{\hbar^2\epsilon}{8} (
 \Gamma \Gamma + R),
\label{path9}
\eqe
where we define $x_N=z$ and $x_0=y$, and $\e=\b/N$.
We decompose $x_k^j$ into a
background and a quantum part, and $S$ into a free and interacting
part
\eq
x_k^j = x_{{\rm bg},k}^j +q_k^j,
\hspace{8mm}  S=S^{(0)}+S^{({\rm int})};
\hspace{8mm} k=1,\ldots, N-1,
\label{path10}
\eqe
where $S^{(0)}=\sum_{k=0}^{N-1} \frac{1}{2\epsilon}
g_{ij}(z) (q_{k+1}-q_k)^i (q_{k+1}-q_k)^j$.
We take the metric in $S^{(0)}$ at $z$ in order to facilitate comparison
with (\ref{path5}), although any other choice should give the same result
as long as one uses the Feynman rules derived below.
In general, $S^{({\rm int})}$
contains in addition to the true interactions also
a pure background piece and
terms linear in $q^j_k$, but to a given order in $\e$ only
a few tadpoles contribute, since the coefficients of these linear terms
will be of order $\e^{1/2}$.
The $N$ factors $g^{1/2}(x_{k+1/2})$ are exponentiated following ref.
\cite{ref2} by
using anti-commuting ghosts $b$ and $c$ and a commuting ghost $a$
\eqa
\sqrt{\det g_{ij} (x_{k+1/2})} & =
& \int  db_{k+1/2}^j dc_{k+1/2}^j da_{k+1/2}^j \nonumber \\ & &
\,\,\,\,\,\,\,
\exp\left( -\frac{\epsilon}{2\beta^2\hbar} g_{ij}(x_{k+1/2})
(b_{k+1/2}^i c_{k+1/2}^j + a_{k+1/2}^i a_{k+1/2}^j) \right).
\label{path11}
\eqae
This formula defines our ghost measure. Introducing modes for the quantum
fluctuations $q$ by the orthonormal transformation
\eq
q^j_k = \sum d^j_m \sqrt{\frac{2}{N}} \sin(\frac{km\pi}{N}) ;
\hspace{2cm} k,m=1,\ldots,N-1
\label{path12}
\eqe
we may change $dx^j_k\rightarrow dq^j_k \rightarrow dd^j_m$. Obviously,
the Jacobian for this transformation is $1$. Next, we couple to
external sources
\eq
S^{({\rm source})}=-\e \sum_{k=0}^{N-1} \left(F_{k+1/2}
\frac{q_{k+1}-q_k}{\epsilon} +
 G_{k+1/2} q_{k+1/2} + \mbox{{\rm sources for ghosts}} \right)
\label{path13}
\eqe
so that we can extract the exact discretized propagators in the usual way.
Completing squares and performing the final integration over
$d^j_m, b^j_{k+1/2}, c^j_{k+1/2}, a^j_{k+1/2}$ leads to $N-1$
factors $g^{-1/2}(z)$ and $N$ factors $g^{1/2}(z)$. Hence
\eq
T=\left(\frac{g(z)}{g(y)}\right)^{1/4}
\exp (-\frac{1}{\hbar} S^{({\rm int})} ) \exp ( -\frac{1}{\hbar}
 S^{({\rm source})}).
\label{path14}
\eqe
{}From $S^{({\rm int})}$ we find the vertices while $S^{({\rm source})}$
yields the discretized propagators. Defining $\dot{x}_{k+1/2}
=(x_{k+1}-x_k)/\epsilon$ and omitting superscripts and a factor
of $\hbar$ for the time being, they come out as follows \cite{ref3}
\eqa
<q_{k+1/2} q_{l+1/2} > & = & -\frac{\epsilon}{4N} (2k+1)(2l+1)
+\frac{\epsilon}{2} (2\min(k,l)+\eta_{k,l}) \nonumber\\
<q_{k+1/2} \dot{q}_{l+1/2} > & = &
-\frac{k+1/2}{N} + \theta_{k,l} \nonumber\\
<\dot{q}_{k+1/2} \dot{q}_{l+1/2} > & = &
-\frac{1}{N\epsilon} + \frac{1}{\epsilon} \delta_{k,l} \nonumber \\
< b_{k+1/2} c_{l+1/2} > & = & -\frac{2}{\epsilon} \delta_{k,l}
\nonumber \\
< a_{k+1/2} a_{l+1/2} > & = & \frac{1}{\epsilon} \delta_{k,l},
\label{path15}
\eqae
where $\eta_{k,l}=1$ if $k\neq l$ but $\eta_{k,l}=1/2$ if $k=l$, while
$\theta_{k,l}=0$ if $k<l$, $\theta_{k,l}=1/2$ if $k=l$ and
$\theta_{k,l}=1$ if $k>l$. We require that $x_{{\rm bg},k}^{j}$ satisfies
the boundary conditions and the
equation of motion of $S^{(0)}$. In the continuum limit this
becomes $x^j_{{\rm bg}}(t)=z^j+(z-y)^j (t/\b)$, while $q^j(t)$
vanishes at the endpoints. In this limit the two-point functions
become (reinstating the superscripts, factors of $\hbar$, and
recalling that $t=\beta\tau$)
\eqa
<q^i(\sigma) q^j(\tau) >  & = & -\beta \hbar g^{ij}(z)
\Delta(\s,\t)
\nonumber \\
<b^i(\sigma) c^j(\tau) > & = & -2 \beta \hbar g^{ij}(z)
 \partial_{\sigma}^2 \Delta(\sigma,\tau)
\nonumber \\
<a^i(\sigma) a^j(\tau) > & = &  \beta \hbar g^{ij}(z)
 \partial_{\sigma}^2 \Delta(\sigma,\tau)
\nonumber \\
\Delta(\sigma,\tau)& = & \sigma(\tau+1)\theta(\sigma-\tau) +
 \tau(\sigma+1) \theta(\tau-\sigma) .
\label{path16}
\eqae
Note that $\Delta(\sigma,\tau)=\Delta_F(\sigma-\tau) + \sigma\tau
+\frac{1}{2}(\sigma+\tau)$, where $\Delta_F(\sigma-\tau)=
\frac{1}{2}(\sigma-\tau)\theta(\sigma-\tau)+
\frac{1}{2}(\tau-\sigma)\theta(\tau-\sigma)$
is the Feynman propagator,
and formally
$\partial_{\sigma}^2\Delta(\sigma,\tau)=\delta(\sigma-\tau)$ while
$\Delta(\sigma,\tau)=0$ at the boundaries. However, {\it the
$\delta(\sigma-\tau)$ is a Kronecker delta} and moreover
{\it the equal-time contractions are unambiguously defined}. Kronecker
delta here means that $\int dx \delta(x) f(x)=f(0)$, even when $f$ is
a product of distributions.
{}From (\ref{path15}) we further find in the continuum limit
\eqa
<q^i(\sigma) \dot{q}^j(\tau)> & = &
-\beta \hbar g^{ij}(z) (\sigma+\theta(\tau-\sigma))
\nonumber \\
<\dot{q}^i(\sigma) \dot{q}^j(\tau) > & = &
-\beta \hbar g^{ij}(z) (1-\delta(\sigma-\tau) ).
\label{path17}
\eqae
All propagators are now proportional to $\beta\hbar$ (this motivated
the normalization of the ghost action in (\ref{path11})), and the
interactions are given by
\eqa
\frac{1}{\hbar} S^{({\rm int})}  & =
& \frac{1}{\beta\hbar} \int_{-1}^0 \left[
\frac{1}{2} g_{ij} (x_{\rm cl}+q)
\left\{
( \dot{x}_{\rm cl} + \dot{q})^i
( \dot{x}_{\rm cl} + \dot{q})^j + b^ic^j + a^ia^j \right\}
-\frac{1}{2} g_{ij}(z) \dot{q}^i \dot{q}^j
\right] d\tau
\nonumber \\
& & - \beta \hbar \int_{-1}^0 \frac{1}{8} (\Gamma \Gamma + R) d\tau.
\label{path18}
\eqae

To compute the configuration space path integral, we note that we must
expand the prefactor $g^{1/4}(z)/g^{1/4}(y)$, which came from the measure,
and evaluate all vacuum graphs with external $x_{\rm bg}$,
using the propagators in (\ref{path16}), (\ref{path17})
and the vertices in (\ref{path18}).
The $q$-independent
part of $S^{({\rm int})}$ does not yield the full $S_{\rm cl}$ of
(\ref{path5}) since $x_{\rm bg}$ is only a solution of the $S^{(0)}$ equation
of motion; rather, tree graphs with
two vertices from
$S^{({\rm int})}$ contribute to order $\b$ as
well. In the one-loop graphs with one vertex $S^{({\rm int})}$ one
finds from the equal-time contractions $<\dot{q}^i \dot{q}^j +
b^ic^j + a^ia^j>$ a contribution proportional to $\int_{-1}^0
\t (\partial_{\s} \partial_{\t} \Delta +\partial^2_{\s} \Delta)_{\s=\t}=
-\frac{1}{2}$ times $(z-y)^k\partial_k g_{ij}(z) g^{ij}(z)$,
which cancels a similar contribution from the non-trivial measure
factor in
(\ref{path14}). There are many other one-loop and two-loop graphs, and
the contribution of each corresponds to a particular term in (\ref{path5}).
In particular, the two-loop graph with one $\dot{q}\dot{q}$, one $\dot{q}q$ and
one $q\dot{q}$ propagator leads to the integral
\eq
\int_{-1}^0 d\s \int_{-1}^0 d\t \partial_{\s} \partial_{\t}
\Delta(\s,\t) \partial_{\s} \Delta(\s,\t) \partial_{\t} \Delta(\s,\t),
\label{18a}
\eqe
where $\partial_{\s} \Delta(\s,\t)=\t+\theta(\s-\t)$ and
$\partial_{\s} \partial_{\t} \Delta(\s,\t)=1-\delta(\s-\t)$, and one
finds agreement with (\ref{path5})
only if one uses (\ref{path1}).
Adding all contributions we have found complete agreement. The non-covariant
vertices $\frac{1}{8}(\Gamma\Gamma + R)$ conspire with the non-covariant
vertices found by expanding $g_{ij}(x)$
in (\ref{path18}) and the measure in (\ref{path14}),
and yield the Einstein invariant
expression (\ref{path5}). The Feynman rules one has to use in this
calculation follow from (\ref{path15}), and they amount to the following.
First, one writes down expressions for all Feynman diagrams using the
propagators given by (\ref{path16}). Adding everything,
all divergencies coming from products of delta functions will
cancel (the ghosts of \cite{ref2} are crucial for this).
The resulting integrals should be worked out using the rules that
delta functions should really be seen as Kronecker deltas and
that $\theta(0)=1/2$. If there are explicit delta functions in the
integrals, one should be careful with partial integrations
and identities like $\int_a^b f'=f(b)-f(a)$, since these are not always
compatible with our Kronecker delta prescription\footnote{The
easiest example is $\int_{-1}^1 dx \delta(x) \theta^n(x)$, which is
$2^{-n}$ according to our prescription, but $1/(n+1)$ if one uses
$\d\theta^n=(\theta^{n+1})'/(n+1)$. The reason the latter is incorrect
is that the discretized derivative $\Delta$ which satisfies
$\sum \Delta f=f(b)-f(a)$ is given by $\Delta f(k+1/2)=(f(k+1)-f(k))/\e$,
but this $\Delta$ does not map the discretized $\theta$ function to the
discretized $\delta$ function in the form in which they appear in
(\ref{path15}). Conversely, we could try to define an operator $\Delta_2$ by
the requirement $\sum \Delta f \Delta g = - \sum \bar{f} \Delta_2
\Delta g + (\bar{f}\Delta g)(b-\e/2)-(\bar{f}\Delta g)(a+\e/2)$,
with $\bar{f}(k+1/2)=(f(k)+f(k+1))/2$,
so that it would
be guaranteed that $\Delta_2$ maps the discretized $\theta$ function
into the discretized $\delta$ function. The problem with this is that
we cannot define $\bar{f}\Delta g$ strictly at the boundary, and if
$\Delta g$ would be a discretized version of $\theta(x-b)$ we would
find that $\lim_{\e\rightarrow 0} (\bar{f}\Delta g)(b-\e/2)=0$, which
would disagree with the continuum result $f(b)/2$.}.
For example, the integral in (\ref{18a}) yields $-1/6$, but if we first
partially integrate the $\partial_{\s}$ in $\partial_{\s} \partial_{\t}
\Delta(\s,\t)$ we find $1/12$.

Finally we discuss the phase space approach. We again decompose $x_k^j$ into
a quantum part $q_k^j$ and a background part $x_{{\rm bg},k}^j$, but
we do not decompose $p^j_{k+1/2}$ (although one could do so). We
couple $p_{k+1/2}$ to external sources $F_{k+1/2}$, and the midpoint
fluctuations $q_{k+1/2}=\frac{1}{2}(q_{k+1}+q_k)$ again to $G_{k+1/2}$.
As free part of the Hamiltonian we take $H^0=\sum_{k=0}^{N-1}
\frac{1}{2} g^{ij}(z) p_{k+1/2,i} p_{k+1/2,j}$. All factors $g^{1/4}$
from $|x>g^{1/2}<x|$ and $<x|p>$ cancel, except at the endpoints, and we find
\eqa
T & = & \left( \frac{g(z)}{g(y)} \right)^{1/4}
 (2\pi\hbar)^{-nN} \exp \left( -\frac{\e}{\hbar} \sum_{k=0}^{N-1}
H^{{\rm int}}_{k+1/2}
 \left( \frac{i\hbar}{\epsilon} \frac{\partial}{\partial F},
  \frac{\hbar}{\epsilon} \frac{\partial}{\partial G} \right) \right)
\int
\prod_{j=1}^n \left( \prod_{k=0}^{N-1} dp_{k+1/2,j}
 \prod_{l=1}^{N-1} dx_l^j \right) \nonumber \\
& &
\! \! \! \! \! \! \! \! \! \! \! \!
\! \! \! \! \! \! \! \! \! \! \! \!
 \! \! \! \! \! \! \!
\times \exp  \sum_{k=0}^{N-1} \left(
-\frac{\epsilon}{2\hbar} g^{ij}(z)
p_{k+\frac{1}{2},i} p_{k+\frac{1}{2},j}
+ \frac{i}{\hbar} p_{k+\frac{1}{2},j} (q_{k+1}^j-q_k^j)
-\frac{i\epsilon}{\hbar} F^j_{k+\frac{1}{2}} p_{k+\frac{1}{2},j}+
\frac{\epsilon}{\hbar} G_{k+\frac{1}{2},j}
q^j_{k+\frac{1}{2}}  \right).
\eqae
We moved the terms $p\dot{x}_{{\rm bg}}$ to $H^{\rm int}$; they give rise
to tadpoles, but to a given order in $\b$ only a few tadpoles contribute,
like in configuration space.
Completing squares and performing the $p$-integrals, we find in the
exponent a term with
\eq
\left( \frac{\epsilon}{2\hbar} \right) \left\{ -i F^j_{k+1/2}
 + \frac{i}{\epsilon} (q^j_{k+1}-q^j_k) \right\}^2.
\eqe
Expanding this term, we regain the discretized configuration space
path integral, multiplied by a factor
\eq
\exp \left( -\frac{\e}{2\hbar} \sum_{k=0}^{N-1} g_{ij}(z)
 F^i_{k+1/2} F^j_{k+1/2} \right).
\eqe
It follows that the discrete $qq$ and $pq$ propagators in phase space
are the same as the $qq$ and $i\dot{q}q$ propagators in configuration
space, but the $pp$ propagator has an extra term
\eq
< p_{k+1/2,i} p_{l+1/2,j} > =
-g_{ii'}(z) g_{jj'}(z)
< \dot{q}_{k+1/2}^{i'} \dot{q}_{l+1/2}^{j'} >
+\frac{\hbar}{\e} g_{ij}(z) \delta_{k,l}
\label{path22}
\eqe
The last term cancels the Kronecker delta present in (\ref{path17}),
so that the $pp$ propagator is non-singular. The continuum limit is
now obvious
\eqa
<p_i(\s) p_j(\t) > &= & \b\hbar g_{ij}(z) \nonumber \\
<q^i(\s) p_j(\t) > & = & -i\b \hbar \delta^{i}_{j} (\s+\theta(\t-\s)).
\eqae
No subtleties like (\ref{path1}) are needed to
evaluate the loops in phase space.

In the discretized phase space approach the boundary conditions
$x_0^j=y^j$ and $x_N^j=z^j$ are taken into account in
$x^j_{{\rm bg},k}$. It is instructive
to compare this with the more conventional continuum approach. There
the propagators for $q(\t)$ and $p(\t)$ form a $2\times 2$ matrix, and
are determined by the equation $F(\s) G(\s,\t)=
\delta(\s-\t) {\bf 1}_{2\times 2}$
and the boundary conditions $q(\s=-1)=q(\s=0)=0$,
where $F$ is the field operator corresponding to $ip\dot{q}-H^{(0)}$,
\eq
F=\left( \begin{array}{cc} 1 & -i\partial_{\s} \\
                           i\partial_{\s} & 0
         \end{array} \right).
\eqe
The solution is
$G(\s,\t)=G_F(\s-\t)+P(\s,\t)$, and Bose symmetry implies
$G^{\a\b}(\s,\t)=G^{\b\a}(\t,\s)$, where $\a=1,2$ denotes $p$ or $q$.
The polynomial $P$ is partly fixed by $FP=0$, while
the boundary conditions on $q(\s)$ fix the rest. Hence,
one should not impose any boundary conditions on $p_j(\s)$, nor
is there any need since the $p$-integrals are all convergent
(in contrast to $q$, $p$ has no zero modes). The
result reads
\eq
G(\s,\t)=
\left(
\begin{array}{cc} 0 & -\frac{i}{2} \epsilon(\s-\t) \\
 \frac{i}{2} \e(\s-\t) & \frac{1}{2} (\s-\t) \theta(\s-\t) +
\s \leftrightarrow \t \end{array} \right) +
\left( \begin{array}{cc} 1 & -i(\t+\frac{1}{2}) \\
 -i(\s+\frac{1}{2}) & \s\t + \frac{1}{2} (\s +\t) \end{array} \right)
\eqe
Given the continuum vertices and propagators in phase space, one can
now again compute the transition element. The vertices are different
because one now expands $g^{ij}(x)$ instead of $g_{ij}(x)$, but
also the propagators are different (no $\delta(\s-\t)$ nor
ghosts). Matthews' theorem asserts
that one should again find (\ref{path5}), and its validity will be
demonstrated elsewhere in more detail \cite{ref3}.

In this article we have found the
correct way to evaluate Feynman diagrams for bosonic non-linear
sigma models in one dimension.
In \cite{bas} an explicit two-loop calculation of the trace anomaly
revealed that the naive Feynman rules together with mode
regularization \cite{ref2} give an incorrect result. We have traced
the origin of this discrepancy to the presence of products of distributions.
These products of distributions were
defined by going back to the origin of the path integral as a
time-discretized expression, and studying the discrete propagators
in (\ref{path15}) and (\ref{path22}). The non-covariant term $\Gamma\Gamma$
in the action (\ref{path18}) can have interesting implications for
field theory in higher dimensions.
In local field theories, it will not contribute if one uses dimensional
regularization. However, it might give rise to additional
finite terms in the action if one makes a non-local field redefinition,
for example when one makes a transformation to go from the axial to
the Coulomb gauge in QCD. Indeed, it has been claimed that such
extra terms appear \cite{schw,chlee}. This,
supersymmetric generalizations, and actual computations
of anomalies, will appear elsewhere \cite{ref3}.

\end{document}